\documentclass[structabstract]{aa}  
\usepackage{epsfig ,graphicx}
\usepackage{txfonts}
\usepackage{natbib}
\bibpunct{(}{)}{;}{a}{}{,} 
\usepackage{natbib,twoopt}
 \usepackage[breaklinks=true]{hyperref} 
 \bibpunct{(}{)}{;}{a}{}{,}    
 \newcommandtwoopt{\citeads}[3][][]{\href{http://adsabs.harvard.edu/abs/#3}%
                                        {\citealp[#1][#2]{#3}}}
 \newcommandtwoopt{\citepads}[3][][]{\href{http://adsabs.harvard.edu/abs/#3}%
                                        {\citep[#1][#2]{#3}}}
 \newcommandtwoopt{\citetads}[3][][]{\href{http://adsabs.harvard.edu/abs/#3}%
                                        {\citet[#1][#2]{#3}}}
 \newcommandtwoopt{\citeyearads}[3][][]%
   {\href{http://adsabs.harvard.edu/abs/#3}{\citeyear[#1][#2]{#3}}}
\begin{document}

   \title{Near-Infrared Imaging Polarimetry of HD142527
    \thanks{Based on observations made with the \textit{VLT} as part of program 089.C-0480(A).}}


   \author{H. Canovas\inst{1,9},
   F. M\'enard\inst{2,9},
   A. Hales\inst{3,4,9},
   A. Jord\'an\inst{5,9},
   M. R. Schreiber\inst{1,9},
   S. Casassus\inst{6,9},
   T. M. Gledhill\inst{7},
   C.  Pinte\inst{8}
   }

   \institute{Departamento de Fisica y Astronom\'ia, Universidad de Valpara\'iso, Valpara\' iso, Chile\\
   \email{hector.canovas@dfa.uv.cl}
   \and
   UMI-FCA, CNRS / INSU France (UMI 3386), and Departamento de Astronom\'ia, Universidad de Chile, Santiago, Chile
    \and
    Atacama Large Millimeter/Submillimeter Array, Joint ALMA Observatory, Alonso de C\'ordova 3107, Vitacura 763-0355, Santiago - Chile
    \and
    National Radio Astronomy Observatory, 520 Edgemont Road, Charlottesville, Virginia, 22903-2475, United States
    \and
    Instituto de Astrof\'isica, Pontificia Universidad
Cat\'olica de Chile, Av.\ Vicu\~na Mackenna  4860, 7820436 Macul,
Santiago, Chile
    \and
    Departamento de Astronom\'ia, Universidad de Chile, Casilla 36-D, Santiago, Chile
    \and
    Science and Technology Research Institute, University of Hertfordshire, College Lane, Hatfield AL10 9AB, UK
    \and
    UJF-Grenoble 1 / CNRS-INSU, Institut de PlanŽtologie et d'Astrophysique de Grenoble (IPAG) UMR 5274, Grenoble, F-38041, France
    \and
    Millenium Nucleus ``Protoplanetary Disks in ALMA Early Science'', Universidad de Chile, Casilla 36-D,Santiago, Chile 
    }

\date{Received May 21, 2013, Accepted YYY, 2013}

    \abstract
     {HD\,142527 is a pre-transition disk with strong evidence for on-going planet formation. Recent observations show a disrupted disk with spiral arms,
     a dust-depleted inner cavity and the possible presence of gas streams driving gas from the outer disk towards the central star.}
   {We aim to derive the morphology of the disk, as well as the distribution and properties of the dust at its surface.}
   {We have obtained polarized differential images of HD\,142527 at $H$ and $Ks$ bands with NaCo at the VLT. Combining these images with classical PSF-subtraction,
   we are able to derive the polarization degree of this disk.}
     {At $H$ band the polarization degree of the disk varies between $10\%$ and $25\%$. 
     This result cannot be reproduced by dust distributions containing highly porous material. The polarization is better matched by distributions of compact particles, with maximum
     sizes  at least up to a few microns, in agreement with previous observations. We also observe two regions of low emission (nulls) in total and in polarized intensity. In particular,
     one of these nulls is at roughly the same
     position as the maximum of the horse-shoe shape observed in sub-millimeter continuum emission ALMA Band-7 (345 GHz) observations. We discuss the possible link between both features.}
 {}

\titlerunning{Near-Infrared Imaging Polarimetry of HD142527}
\authorrunning{Canovas et al.}
\maketitle
\maketitle

\section{Introduction}
Transition disks are a particular case of protoplanetary disks. Their identifying characteristic is a decrement
of near/mid infrared flux when compared to the median of the classical T Tauri stars (CTTs) in the Taurus
cloud \citep[e.g.][and references therein]{Williams_2011}. This observational feature is explained
by a drop in the optical depth ($\tau_{\nu}$). This drop cab be caused by a decrement in opacity or by a drop of dust density
(i.e. a gap/cavity). Binarity, planet formation, dust growth and photoevaporation are the proposed mechanisms to explain
the presence of these ``optical depth'' holes \citep[e.g.][]{2011ApJ...738..131D, 2004A&A...421.1075D, 2012MNRAS.422.1880O}. Each one of them can be identified
by measuring observables such as accretion rates, SED shapes or disk masses \citep{2007MNRAS.378..369N, 2012ApJ...750..157C, 2012ApJ...749...79R}.
Recent studies combining observations at different wavelengths have tried to identify the dominant mechanism
in a sample of transition disks \citep[among others,][]{2011ApJ...742L...5A, 2012ApJ...753...59M, 2012ApJ...747..136I, 2012ApJ...758L..19H,  2012ApJ...750..157C, 2013Natur.493..191C}.
In any case, it is evident that transition disks represent a very important phase in the evolution of the
protoplanetary disk, and that understanding them is fundamental to explain planet formation.

In this paper we focus on HD\,142527, a remarkable transitional disk around a Herbig Ae star. This disk features an inner
disk, a dust-depleted gap and an outer disk, which places it in the category of pre-transition disks \citep{2007ApJ...670L.135E}.
\citet{2011A&A...528A..91V} finds that the inner disk can be fitted assuming a radius of $\le30$ AUs, the gap extends up to
$\approx130$ AUs, and that the scale height at the inner wall of the outer disk must be extremely high ($\approx60$AUs).
HD\,142527 has been imaged at near-infrared (NIR) by \citet{2006ApJ...636L.153F}, \citet{2012ApJ...754L..31C} and \citet{2012A&A...546A..24R}, showing the presence
of several spiral features on the innermost parts of the outer disk. The disk is believed to be inclined by $\approx 20^{\circ}$,
as suggested by mid-infrared (MIR) and NIR observations, and to be rotating in clock-wise direction, as suggested by the
orientation of its spiral features. More recently, \citet{2013Natur.493..191C} confirms the horse-shoe shape of the continuum
emission in this system first noted by \citet{2008Ap&SS.313..101O}, and finds two possible dense gas streams transporting
material from the outer disk towards the central star. These gas streams are predicted by current models of giant-planet
formation \citep{2011ApJ...738..131D}, so they could be a direct signpost of planet formation. \citet{2012ApJ...753L..38B}
discusses the presence of a possible close-in binary companion, although its existence needs confirmation.
\newline

We present polarized differential imaging (PDI) observations of HD\,142527 at $H$ and $Ks$ bands obtained with the
NaCo instrument at the Very Large Telescope (VLT). By means of polarimetry, it is easy to remove most of the (unpolarized)
stellar light to image the circumstellar environments in \textit{scattered} light. Nowadays PDI is becoming a standard technique
to directly image protoplanetary disks, as proven by the increasing amount of systems observed with polarimeters
at 8-m class telescopes \citep[among others][]{2013ApJ...766L...2Q, 2012ApJ...760L..26M, 2012PASJ...64..124T, 2012ApJ...758L..19H, 2012ApJ...753..153K}. 
In most of these studies, only the polarized intensity ($P_I$) image of the disk is presented. Although $P_I$ images directly
show the disk's surface in scattered light, the information contained in these images is limited and cannot be used to constrain the
properties of the dust scattering particles, unless $P_I$ images at different wavelengths are combined. This is because $P_I$ is the
product of two quantities: the total intensity and the \textit{polarization degree}. A classical example of the degeneracies left when
analyzing polarized intensity images alone is the discussion about the ``hole" detected in polarized intensity in AB Aurigae
\citep[see][]{2008ApJ...679.1574O, 2009ApJ...707L.132P}. Using the polarization degree it is possible to break some of these
degeneracies and to put constrains on the size, composition and/or shape of the dust particles
if enough wavelength coverage is  available \citep{2012ApJ...750..161D, 2012A&A...537A..75M, 2010A&A...518A..63M}. In this work we
combine PDI with classical PSF-subtraction to derive the polarization degree in HD\,142527.
The observations and data reduction are described in Sect. 2 and Sect. 3, respectively. Our results are presented in section 4.
The discussion and final conclusions are given in Sections 5 and 6, respectively.


\section{Observations}

The observations were performed in service mode during July and August 
2012, with NaCo \citep{2003SPIE.4841..944L,2003SPIE.4839..140R}
at the VLT/UT\,4. NaCo is a very flexible 
instrument that combines an Adaptive Optics (AO)-system
with a NIR camera. We used NaCo in its polarimetric mode. In this setup,
a half-wave plate (HWP) rotates the polarization plane of the light
by  $\phi = 2\times \theta$, where $\theta$ is the angle measured East of North.
A Wollaston
prism\footnote{A Wollaston prism separates the light by means of 
birefringence. The component
of the light that vibrates parallel to the optical axis of the 
prism is affected by the $extraordinary$ refraction index, $n_e$,
while the perpendicular component of the light is affected by the $ordinary$  refraction index, $n_o$.}
placed after the HWP separates the incoming light into two orthogonally polarized beams, which
are projected on different regions of the camera. The images generated 
by these two beams are hereafter named as $I_{o}$ (for the $ordinary$ beam), and $I_{e}$ (for the $extra-ordinary$ beam).
Each frame recorded by the camera contains these two simultaneous images with orthogonal polarization
states. The separation between the two images is fixed to 3.3" in the 
y-direction of the detector, and remains constant during the whole observing run. 
A field mask prevents beam-overlapping. The pixel size of the camera was set to 0.027"/px, the readout mode to $Double\_RdRstRd$ and
detector mode to $HighDynamic$. The North-South direction on the sky coincides
with the vertical axis on the detector and with the zero position of the fast axis of the HWP (i.e., $\theta=0^{\circ}$).

\subsection{Data Acquisition}

HD\,142527 and the comparison star HD\,161743  were observed
in $H$ and $K_s$ bands. A brief description of both objects is given
in Table~\ref{tab:targets}. Four datasets taken with the HWP rotated
by $\theta = 0^{\circ}$, $22.5^{\circ}$, $45^{\circ}$ and $67.5^{\circ}$
were recorded during each observation. In what follows ``dataset"
always refers to images recorded under the same HWP
position angle. An observing Log is given in Table~\ref{tab:observations}.
HD\,142527 was observed using exposure times of 0.4, 1 and 5 seconds 
in the $H$ band and  0.4 and 4 seconds in the $K_s$ band. 
Three subsets of images per dataset (Ditpos) recorded 
at different regions on the detector
were obtained to minimize the effect of bad pixels on the final images.
The comparison star HD\,161743 was observed in $H$ band with 
exposure times of 3, 10 and 15 seconds, and in $K_s$ band with 
exposure times of 10 and 15 seconds. Only one subset of images
per HWP angle was recorded in these measurements.
All images except the 0.4s H-band observations of HD\,142527 
were overexposed on the central star. The saturated region reaches a maximum size of radius
$r_{sat}\approx 0.32"$ in the 5s $H$-band images.

Together with the science observations a set of dark frames with the same exposure
times than the science images were acquired. Dome and sky flats were taken
with and without polarization optics, respectively. Different exposure times
ranging from 0.2 to 30 seconds were used during the flat acquisition.

\begin{table}
		\centering
				\caption{Description of HD 142527 and HD 161743.}
			\begin{tabular}{ c  c  c  c  c  c }
\hline\hline
Target			&	RA				&	Dec				&	$\mathrm{m}_H$		&	$\mathrm{m}_{K_s}$		& Spectral		\\
HD				&	[hh:mm:ss]		&	[dd:mm:ss]		&						&							& type			\\
\hline
142527			&	15:56:41.89		&	-42:19:23.27		&	5.71\tablefootmark{1}	&	4.98\tablefootmark{1}		&	F7IIIe\tablefootmark{2}	\\
161743			&	17:48:57.92		&	-38:07:07.48		&	7.57\tablefootmark{1}	&	7.57\tablefootmark{1}		&	B9IV\tablefootmark{3}		\\
\hline
			\end{tabular}
\tablefoottext{1}{\citet{2003yCat.2246....0C}},\tablefoottext{2}{\citet{2005A&A...437..189V}},\tablefoottext{3}{\citet{1982mcts.book.....H}}
	\label{tab:targets}
\end{table}

\begin{table}
		\centering
				\caption{Observing Log of the observations. The number of images (Ndist$\times$Ditpos) is
				given $per$ HWP position angle.}
			\begin{tabular}{ c  c  c  c  c  }
\hline\hline
Target			&	Band	&	Exp Time	&	Ndits$\times$DitPos			&	Date			\\
				&			&		[s]		&							&	[dd/mm/yy]		\\
\hline
HD\,142527			&	H		&		5		&	7x3						&	19/07/2012		\\
				&			&		1		&	32x3						&	"				\\
				&			&		0.4		&	4x3						&	"				\\
				&	Ks		&		4		&	15x3					&	25/08/2012		\\
				&			&		0.4		&	4x3						&	24/08/2012		\\
				&			&		0.4		&	4x3						&	11/08/2012		\\
HD\,161743		&	H		&		15		&	1x1						&	19/07/2012		\\
				&			&		10		&	2x1						&	"				\\
				&			&		3		&	5x1						&	"				\\
				&	Ks		&		15		&	2x1						&	25/08/2012		\\
				&			&		10		&	2x1						&	"				\\
				&			&		10		&	2x1						&	11/08/2012		\\				
\hline
			\end{tabular}
	\label{tab:observations}
\end{table}

\section{Data Analysis}

Polarized light can be described by means of the \textit{Stokes} parameters $I$, $Q$, $U$ and $V$.
Stokes $I$ represents the total intensity. Stokes $Q$ and
$U$ describe linearly polarized light, while Stokes $V$ describes circularly polarized light.
In this work we focus on measuring linearly polarized light, so we do not consider the
Stokes $V$ parameter. The polarized intensity ($P_{I}$) is described
by $P_{I} = \sqrt{Q^2 + U^2}$, while the degree of polarization ($P$) is computed as $P = P_{I}/I$.
The polarization angle ($P_\theta$), which indicates the plane of vibration of the electric field
associated to the light, is computed as $P_{\theta} = \frac{1}{2}arctan (U/Q)$.

There is no standard pipeline to reduce polarimetric NaCo data.
We used our own routines written in IDL. In the following subsections
we describe our data processing method.

\subsection{Basic Reduction}

The individual dark frames were median-combined to generate a set of master dark for each exposure time. 
Master flats were produced in the same way from the individual sky flats (taken without the polarization optics). 
The different master flats were visually inspected to discard those affected by light gradients and artifacts.
Each science image was master dark subtracted and divided by the normalized master flat.
The effect of hot and bad pixels was reduced by sigma-clipping and median-combining
the different subsets of images.
The performance of the AO-system was stable during the whole run,
and no images were discarded due to poor Strehl ratios. Furthermore, given
the discrete amount of images obtained per dataset (see Table~\ref{tab:observations})
we use all the observations to increase the signal to noise ratio (S/N) of the final images.

Image alignment is a critical issue in polarimetric imaging 
\citep[e.g.,][]{2004A&A...415..671A,2011A&A...531A.102C},
and must be done with sub-pixel accuracy. We performed several tests combining
different aligning procedures (shift-and-add, cross-correlation with different templates)
to check which combination produced the best alignment. We conclude that separately
aligning each dataset by means of a cross-correlation algorithm (accuracy
of 1/5 of a pixel) provides the best results for our images. The aligning procedure
works as follows. First, all images are centered with a classical shift-and-add
method to generate a template. Second, the $I_o$ of each frame is centered with 
respect to this template by means of a cross-correlation. This process is repeated 
to center the $I_e$ using the $I_o$ image as a reference. 
This process is applied in the same way to each pair of $I_o$
and $I_e$ images.

The sky background is computed as the median of two different sky regions
($0.29"\times0.29"$ each), and then subtracted from the individual $I_o$ and
$I_e$ images. Those sky regions are located at 4" away from the star to ensure
that the contribution from the stellar Point Spread Function (PSF) is negligible. 

\subsection{Polarimetry}

Once the images are aligned and sky-subtracted, they are combined to extract
the Stokes parameters $I$, $Q$ and $U$. 
There are two standard approaches to do this in imaging polarimetry: 
the ``difference" and ``ratio" methods
\citep[see among others, ][]{1996SoPh..164..243K, 2001ApJ...553L.189K, 2009ApJ...701..804H, 2011ApJ...738...23Q, 2011A&A...531A.102C,2012A&A...543A..70C}.
The difference method is less sensitive to PSF variations and more sensitive
to flat-field variations, while the ratio method behaves in the opposite way
\citep[for a detailed discussion on the topic see][]{2011A&A...531A.102C}.
Both methods were tested in our analysis, concluding that the 
difference-method delivers results with a higher signal-to-noise (S/N) ratio.
In what follows we describe how the polarized light is extracted from the aligned images
by means of the difference method.

The individual Stokes Q images are produced from each pair of $I_o$ and $I_e$ images taken with the HWP rotated at
$\theta = 0^{\circ}, 45^{\circ}$ (i.e., the polarization plane is rotated by $\phi =  2\times\theta = 0^{\circ}, 90^{\circ}$):
\begin{eqnarray}
+Q_{ind} 	& = & I^{\theta}_{o} - I^{\theta}_{e}\bigr|_{\theta = 0^{\circ}}  \\
-Q_{ind}	& = & I^{\theta}_{o} - I^{\theta}_{e}\bigr|_{\theta = 45^{\circ}}
\label{eq:eqQQ}
\end{eqnarray}
Similarly, the individual Stokes U images are obtained from:
\begin{eqnarray}
+U_{ind} 	& = & I^{\theta}_{o} - I^{\theta}_{e}\bigr|_{\theta = 22.5^{\circ}}  \\
-U _{ind}& = & I^{\theta}_{o} - I^{\theta}_{e}\bigr|_{\theta = 67.5^{\circ}}
\label{eq:eqUU}
\end{eqnarray}
Each $+Q/\mathrm{-}Q$ and $+U/\mathrm{-}U$
image is corrected of instrumental polarization.
Individual intensity images at each HWP position ($I^\theta$) are generated by adding the $I_{o}$ and  $I_{e}$ images:
\begin{equation}
I^{\theta}_{ind}  =  I^{\theta}_{o} + I^{\theta}_{e}\bigr|_{\theta = 0^{\circ},22.5^{\circ},45^{\circ},67.5^{\circ}}  \\
\label{eq:eqII}
\end{equation}

\subsubsection{Instrumental Polarization}
\begin{table}
		\centering
				\caption{Average polarization degree within $ring_{IP}$ at $H$-band.}
			\begin{tabular}{ c  c  c  c  c  c c}
\hline\hline
Target				&	Elevation	&	Azimuth		&	Q/I		&	-Q/I		&	U/I		&	-U/I		\\
	HD				&	$[^\circ]$	&	$[^\circ]$	&	$[\%]$	&	$[\%]$	&	$[\%]$	&	$[\%]$	\\
\hline
142527 $(1)$	&	48.277		&	53.162		&	2.29		&	-2.17	&	-0.89	&	0.95		\\
142527 $(5)$	&	50.693		&	52.405		&	1.88		&	-1.91	&	-0.86	&	0.81		\\
161743 $(15)$	&	56.271		&	57.566		&	1.81		&	-1.49	&	-0.40	&	0.72		\\
161743 $(10)$	&	74.138		&	29.416		&	0.65		&	-0.19	&	-0.54	&	0.91		\\
161743 $(3)$	&	57.641		&	56.958		&	2.03		&	-1.42	&	-0.42	&	0.93		\\

\hline
			\end{tabular}
		\tablefoot{Exposure times are shown inside parenthesis, in seconds. Elevation and Azimuth are given at the beginning of the observation.}
	\label{tab:ip}
\end{table}

The instrumental polarization (IP) is usually corrected (in imaging polarimetry) by subtracting the polarized light
measured at the position of the central star. This is equivalent to assuming that the central resolution element is unpolarized. 
This method, previously used by several authors \citep[e.g.][]{2008PASP..120..555P,2011ApJ...738...23Q,2013ApJ...766L...2Q,2011A&A...531A.102C,2012A&A...543A..70C},
efficiently removes any polarized signal produced by the telescope $+$ instrument system as well as the interstellar polarization,
but it may also remove the contribution from the inner disk in HD\,142527 to the polarized light. On the other
hand, the central star of HD\,142527 saturates in almost all our observations inside the inner $r_{sat}\le0.32"$ from the star.
Because of this, we choose to compute the IP in a ring ($ring_{IP}$) of inner radius $r_{i} = 0.35"$ and outer radius of $r_{o} = 0.5"$,
centered on the star. This ring falls inside the huge, dust-depleted, gap of HD\,142527 \citep{2012A&A...546A..24R,2012ApJ...754L..31C,2013Natur.493..191C}.
The average polarization inside $ring_{IP}$ is computed and subtracted from each individual  $+Q/\mathrm{-}Q$ and $+U/\mathrm{-}U$ image.
The average values of the polarization inside $ring_{IP}$ for the $H$-band observations are shown in Table~\ref{tab:ip}.
The IP is extremely sensitive to the telescope orientation, as reflected by our results. 
The interstellar polarization can be estimated as $P_{Int} \le 3 A_{v}$ \citep[see][and references therein]{2003ARA&A..41..241D},
where $A_{v}$ is the total extinction. Using the $A_{v} = 0.6$ derived by \citet{2011A&A...528A..91V}, we obtain a maximum interstellar polarization
of $P_{Int} = 1.8\%$. On the other hand, the instrumental polarization of NaCo can reach a maximum value of
$4\%$ \citep{2011A&A...525A.130W}. Therefore we conclude that inside the gap where we estimate IP, the contribution
from $P_{Int}$ and the instrumental polarization dominates over any polarization signal produced by scattering by dust
(see Table~\ref{tab:ip}, all the values are below $2.3\%$.)
\begin{figure*}[H!]
  \center
   \includegraphics[width = 1\linewidth,trim = 15 40 55 0]{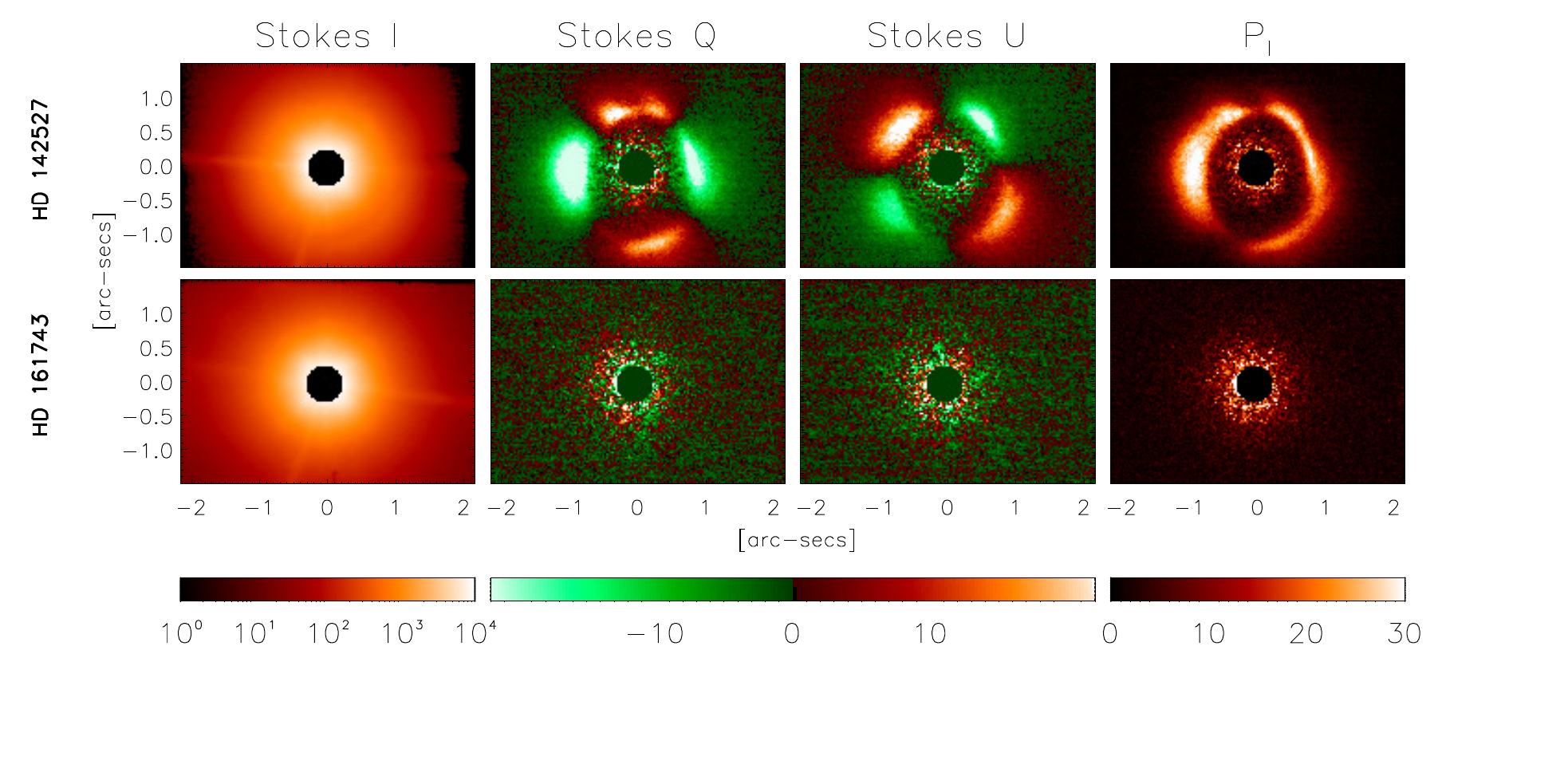}
   \caption{Processed images of HD\,142527 (top row) and HD\,161743 (bottom row) in H band. From left to right: Intensity
   image ($I$) in logarithmic scale, Stokes $Q$, $U$ and $P_{I}$ images in linear scale. For comparison purposes, HD\,161743 has been scaled by an arbitrary factor.
   Units are given in counts. The area corresponding to the saturated pixels ($r_{sat}\le0.32"$) in HD\,142527 has being masked out in all images. The polarized images
   of HD\,142527 show a complex structure, while the comparison star only shows remnant noise. In all images North is up and East is left. This applies to all the figures on this paper.}
  \label{fig:hd14_fig1}
\end{figure*}

\subsubsection{Final Images}
The individual (IP-corrected) $+Q$ images are centered with a cross correlation algorithm
using the first $+Q$ image of the dataset as the centering-template. Once aligned, they
are median-combined to produce a final $+Q$ image. This process is repeated with the
individual $-Q$, $+U$, $\mathrm{-}U$ and $I^{\theta}_{ind}$ images. An intensity
image per dataset ($I^{\theta}$) is created by median-combining the $I^{\theta}_{ind}$.
The final Stokes $Q$ and $U$ images are generated from the median-combined
images as: 
\begin{eqnarray}
Q	& = & 0.5\left(+Q - (-Q) \right) \\
U	& = & 0.5\left(+U - (-U) \right)
\label{eq:eqQU}
\end{eqnarray}
The polarized intensity ($P_I$) and the polarization angle ($P_{\Theta}$)
are computed from these images:
\begin{eqnarray}
P_{I} &=& \sqrt{Q^2 + U^2} \\
P_{\Theta} &=& \frac{1}{2}atan \left(\frac{U}{Q} \right).
\label{eq:Ps}
\end{eqnarray}
\citet{2011A&A...525A.130W} detected and offset of $13.2^{\circ}$ in the HWP of NaCo
that is included in our calculations of $P_{\Theta}$.
The final intensity image ($I$) is produced by averaging the $I^{\theta}$ images
generated for each dataset:
\begin{equation}
I	 =  0.25\left(I^{0^{\circ}} + I^{22.5^{\circ}} + I^{45^{\circ}} + I^{67.5^{\circ}} \right),
\label{eq:eqI}
\end{equation}
where the superscript indicates the $\theta$ angle. Finally, the images are normalized to
1 second exposure time. The $I$, $Q$, $U$ and $P_{I}$ images of HD\,142527 (5s exposure time)
and HD\,161743 (5s exposure time) at $H$-band are shown in Fig.~\ref{fig:hd14_fig1}.

We note that \citet{2013ApJ...766L...2Q}  \citep[see also][]{2006A&A...452..657S} use the radial Stokes parameters $Q_r$ and $U_r$
to produce a polarized intensity image with reduced levels of noise (we refer the reader to that paper
for a detailed explanation of these coefficients). As noted by these authors, this approach is valid
when the disk is nearly face-on. We have reduced our images with the method described by
\citet{2013ApJ...766L...2Q} (and Avenhaus et al. 2013, in preparation), concluding that the use of $Q_r$ and $U_r$ indeed delivers
significantly lower levels of noise in the centermost regions of the images. However,
at larger distances from the image center ($r\ge 0.4"$), the differences between both methods are
within the error bars. Given that we do not consider the centermost regions (r$\le0.3"$) of the images in this work
(due to the saturation of our images), and that the inclination of HD\,142527 is best fit by $\approx20^{\circ}$,
we prefer to use the ``classical" polarized intensity as described in Eq.~\ref{eq:Ps} of this paper.

\subsection{PSF-subtraction}
An important part of the information contained in the polarized light 
is encoded in the degree of polarization ($P = P_{I}/I$). This quantity directly traces 
the size and properties of the dust particles that scatter the light, polarizing it \citep[e.g.][]{2007A&A...470..377V,2012A&A...537A..75M}.
However, the $I$ image derived in the 
previous subsection contains the contribution of the star $+$ disk. Therefore, it is necessary
to obtain an intensity image of the disk ($I_\mathrm{disk}$) alone to derive the
degree of polarization ($P$) of the disk. To that end, we have used the observations of
the comparison star HD\,161743 ($I$ images) to perform PSF-subtraction to the HD\,142527
$I$ images. In what follows, we describe our PSF-subtraction process. 

Before starting the process we check for the presence of ghosts and gradients by subtracting from each image
an azimuthally averaged version of itself. We do this with both the HD\,142527 and the comparison star. By doing
this we have identified several static ghosts (i.e., ghosts
that remain at fixed positions over the detector) that could be mistakenly identified as bright clumps on the disk
of HD\,142527. This is especially important in our $Ks$-band observations, where we find at least twice as many
ghosts/artifacts than in our $H$-band images. Most of these artifacts are unpolarized, since they do not appear in
the corresponding $P_{I}$ images. We also conclude that the images with longer exposure times
are strongly affected by spatial gradients. This can be explained by the lower performance of the AO when observing
at longer exposure times. Because of this, we remove from the analysis the images
taken with the longer exposure times.

\begin{figure}
  \center
   \includegraphics[width = 1\linewidth,trim = 30 20 25 20]{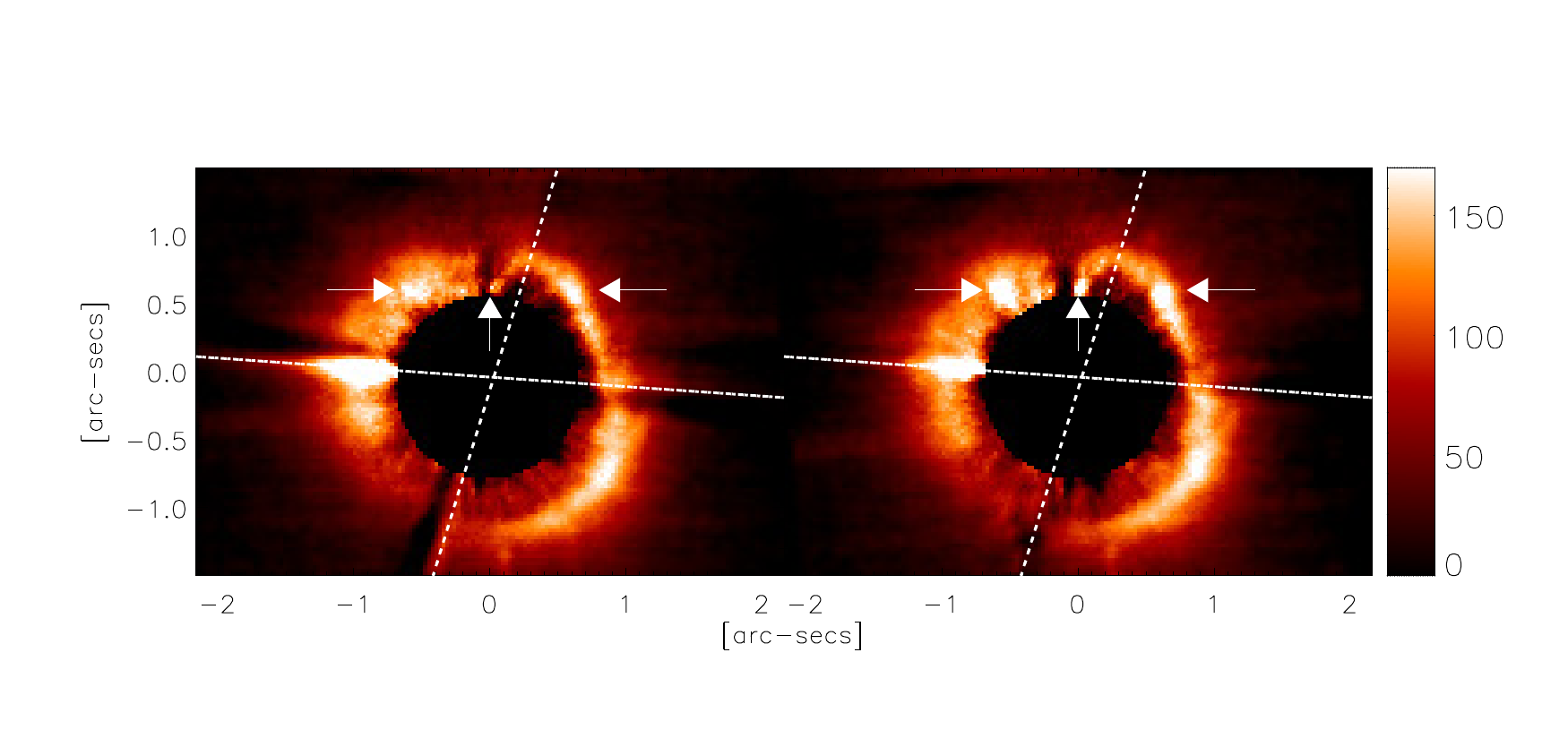}
   \caption{$I_\mathrm{disk}$ images at H-band. Left: without rotating the flux-scaled HD\,161743 image.
   Right: rotating the HD\,161743 image to match the spider's pattern (indicated by dashed-lines) in the HD\,142527 image.
   The innermost ($r \le 0.67"$) regions are masked out to remove artifacts. The white arrows point to
   previously identified ghosts, that are enhanced when rotating the PSF to correct for the spiders. Bar units are given in counts.
   The bright path to the East, right over the doted spider line, is an artifact of the PSF-subtraction process.}
  \label{fig:hd142527_fig2}
\end{figure}

The scaling coefficient $\alpha(r)$ used to flux-scale the
HD\,161743 images was computed as the ratio between HD\,142527 and HD\,161743 at a given position.
$\alpha(r)$ is computed at radii ranging from $r=0.37"$ from the central star (outside of the saturated area)
to $r=0.67"$ (before the outer disk's emission becomes significant). The final scaling coefficient ($\alpha_f$) is
computed as the average of the different $\alpha(r)$. The standard deviation of this coefficient is more than one order
of magnitude smaller than its value, indicating that the radial profiles of HD\,142527 and HD\, 161743 are very similar within
the $[0.37" - 0.67"]$ region from the central star. However, we note that the radial profiles of both stars at larger distances from
the center ($r \ge1.5"$) are fairly different, making impossible a uniform PSF-subtraction over the whole field of view 
\citep[see also][who noticed the same when performing PSF-subtraction for this target at $L$-band]{2012A&A...546A..24R}.
Once the comparison PSFs are flux-scaled, they are aligned and subtracted to the HD\,142527 images. We do not rotate the HD\,161743
(as usually done to match the spider's pattern in both images) to minimize the effect of the static artifacts
(see Fig.~\ref{fig:hd142527_fig2}). Due to the uncertainties of the entire process, we consider our results as lower
limits of the flux of the disk. We use the sky regions of the PSF-subtracted image to give an estimation
of the upper limit of the disk's emission. Computing the mean in 8 different sky regions of over image
(using boxes of $5\times5$ pixels) we obtain a value of $\approx7$ counts for the background. We therefore
use this value as a 1-$\sigma$ estimation. We construct the upper limit to the brightness of the disk
as 3 times this quantity.

\section{Results}
 
\begin{figure}[ht]
  \center
   \includegraphics[width = 1\linewidth,trim = 10 0 20 40]{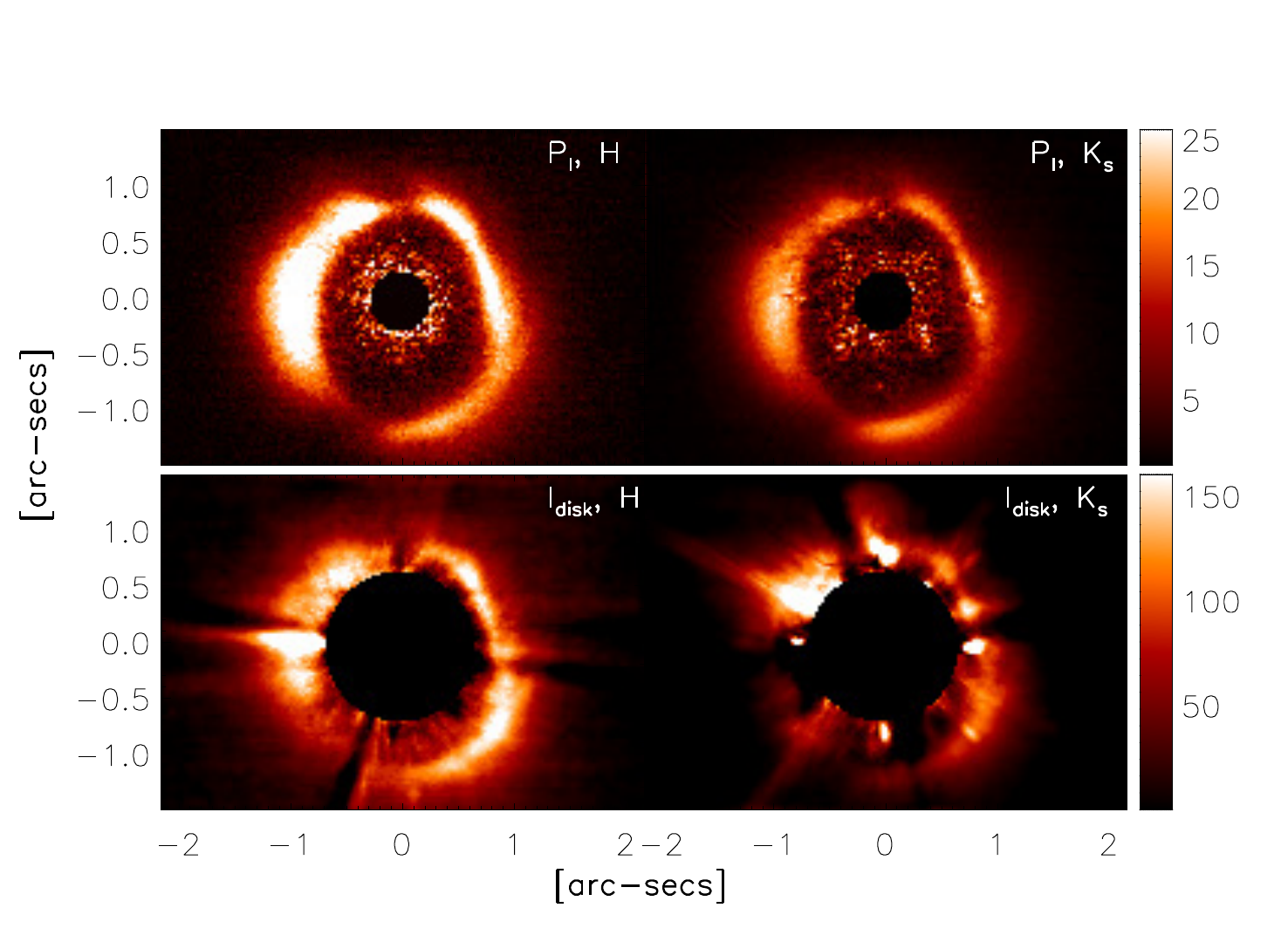}
   \caption{$P_{I}$ (top row) and $I_\mathrm{disk}$ (bottom row) images at $H$-band (left column) and $Ks$-band (right column)
   of HD\,142527. Masked area in the $P_{I}$ images cover the saturate region, while in the PSF-subtracted images cover
   the artifact-dominated regions. The $P_I$ are plotted with the same scale, to enhance differences/similarities. The same
   is done with the $I_\mathrm{disk}$ images. The bright patch in the $I_{\mathrm{disk}}$ image at North-East direction in $Ks$
   band is an artifact due to the PSF-subtraction (as it is the bright path on the East direction in the $I_{\mathrm{disk}}$ image at
   $H$ band, see also the caption in Fig.~\ref{fig:hd142527_fig2}). Color bar units are given in counts.}
  \label{fig:hd142527_fig3}
\end{figure}

The images taken with the shortest exposure time (0.4 s) at $H$ and $Ks$ band are disregarded from this analysis
due to their poor S/N ratios. Because of this, we focus our analysis in the 1s images at $H$-band, and 4s images at $Ks$ band.
Fig.~\ref{fig:hd142527_fig3} shows the $P_{I}$ (top row) and $I_\mathrm{disk}$ (bottom row) images at $H$ (left column)
and $Ks$ band (right column). The two $P_{I}$ images are plotted with the same color scale, as it is done with the $I_\mathrm{disk}$
images in the bottom row. All the bright clumps in the $I_\mathrm{disk}$ image at $K$s band are caused by instrumental, unpolarized, artifacts. 
The overall disk structure recovered from the PSF-subtracted images matches well previous images at $Ks$-band
\citep{2006ApJ...636L.153F,2012ApJ...754L..31C} and $L$-band \citep{2012A&A...546A..24R}. The polarized signal inside the gap
is within $3\sigma_{P_I}$ of the sky background. We estimate $\sigma_{P_I}$ of the background
by computing the median of the standard deviation in 4 sky regions ($5\times5$ px each) of the $P_I$ image.
There is a marginal detection of the spiral feature (PA $\approx 260^{\circ}$) labelled as ``2" in Fig.~2 by \citet{2012ApJ...754L..31C}. Both
the PSF-subtracted and the $P_I$ images at $H$ and $K$s bands show the presence of two nulls or gaps at position angles
of PA: [$340^{\circ}$ to $10^{\circ}]$ (northern null) and PA: [$130^{\circ}$ to $165^{\circ}]$ (southern null). 
Inside the cavity, the best detection limit for a point source in the intensity image is $\Delta m_H = 5.5$ magnitudes at $0.6"$ from the central star.
In polarized intensity, the best $3\sigma$ limit is 13.5 $\mathrm{mags/arcsec^{2}}$ at the same position. With these limits we do not detect the HCO+ streamers claimed by
\citet{2013Natur.493..191C} and cannot verify the presence of the putative companion claimed by \citet{2012ApJ...753L..38B}.

\subsection{Brightness asymmetries and color of the $P_I$ images}
\begin{figure}[ht]
  \center
   \includegraphics[width = 1\linewidth,trim = 35 0 60 40]{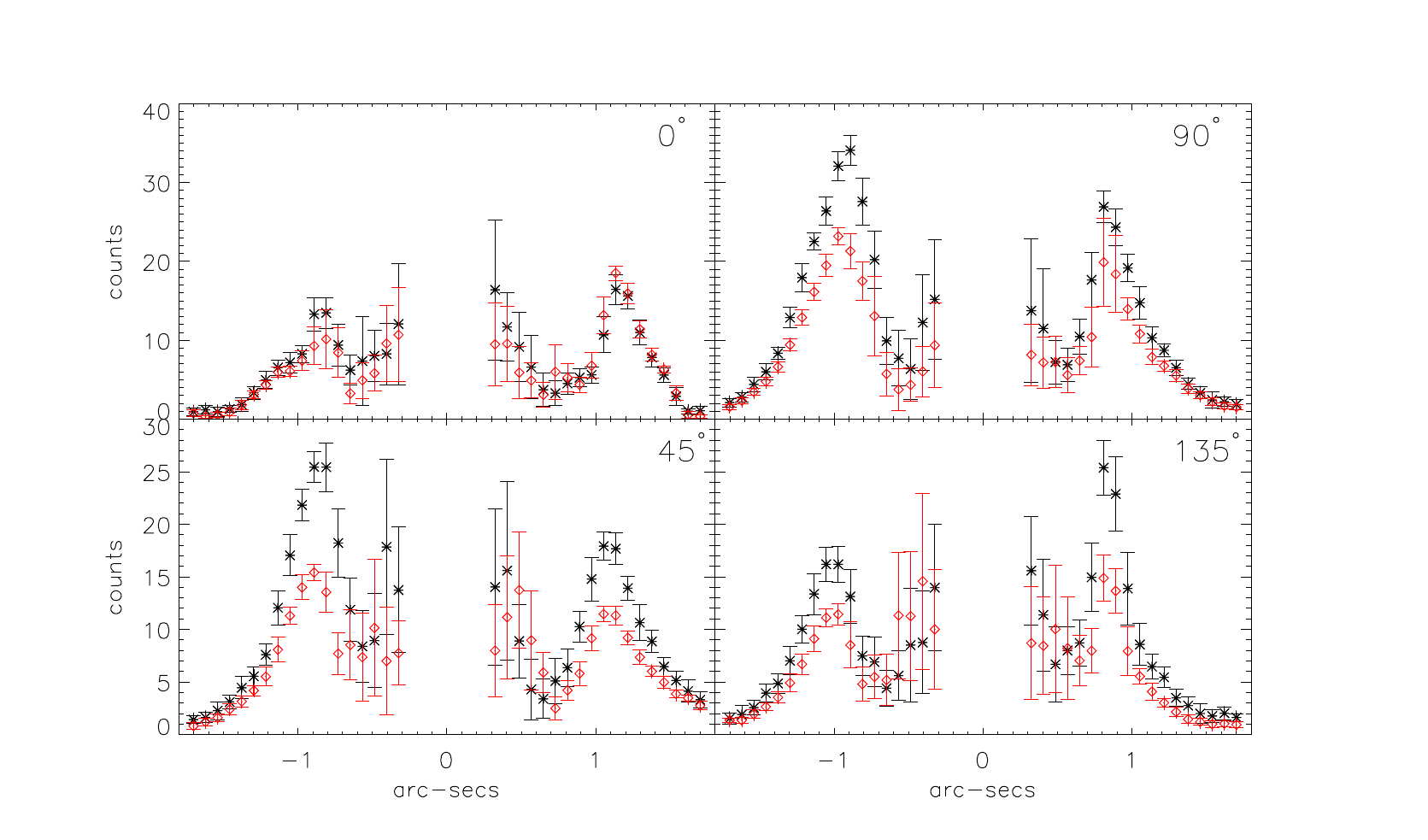}
   \caption{$P_{I}$ radial cuts: $H$ and $Ks$-band cuts are plotted in black asterisks and red diamonds, respectively. Innermost regions have been masked
  out due to their significantly higher noise. Each point represents the mean value of the $P_I$ image at a given distance from the center, within a squared piece
  of $3\times3$px ($0.08"\times0.08"$). The error bars are the standard deviation inside each piece. The number at the upper-right corner in each plot indicates the
  PA (measured East of North) of the cut. The plot at $0^{\circ}$ shows a cut from North (top) to South (bottom). On the other plots, the radial cuts start from the
  eastern side (left-side) of the images to the western side (right-side) of the images.}
  \label{fig:hd142527_fig4}
\end{figure}

\begin{figure}[ht]
  \center
   \includegraphics[width = 1\linewidth,trim = 15 0 5 0]{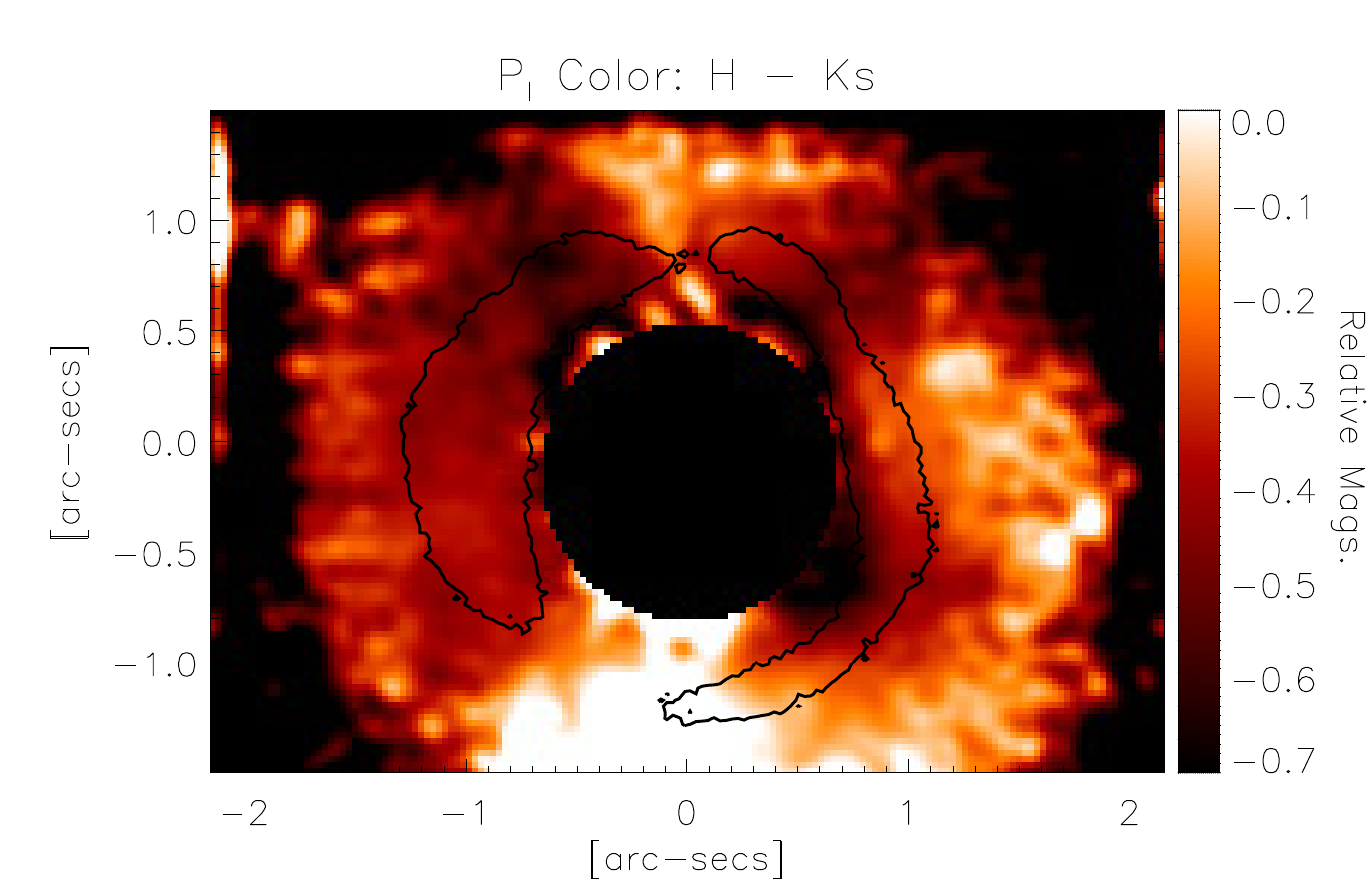}
   \caption{$P_{I}$ color. Inner regions have been masked out to remove artifacts. The black contours outline the
   position of the disk seen in polarized intensity at $H$-band (Fig.~\ref{fig:hd142527_fig3}, top left). 
   The image has been smoothed with a 4-px Gaussian kernel. The two nulls have more neutral colors.}
  \label{fig:hd142527_fig5}
\end{figure}

The Eastern side of the disk is more extended than the Western side in the $P_I$ images at both $H$ and $Ks$ bands. 
The Eastern side is also brighter at $H$-band, while both sides have similar peak
brightness at $Ks$ band, as shown in Fig.~\ref{fig:hd142527_fig3}. Furthermore, the disk shows strong asymmetries along 
its inner rim. To further test this asymmetry we have generated radial cuts on the $P_I$ images at different orientations
instead of computing the radial brightness distribution along the major or minor axis of the disk, as usually done.
The results are shown in Fig.~\ref{fig:hd142527_fig4}. The $H$ and $Ks$ band results are represented by black
stars and red diamonds, respectively. Both images are very similar \textit{only} along the vertical direction ($0^{\circ}$ plot), which is
the major axis of the disk \citep{2012ApJ...754L..31C}. The other plots clearly show that the disk's asymmetries between the East and
West sides in $P_I$ are more accentuated in the $H$-band than in the $Ks$ band.
We show the polarized-color of the disk in Fig.~\ref{fig:hd142527_fig5}). This image has been smoothed with a Gaussian filter of 4-px kernel to reduce
the noise and focus on the large-scale structures. In that figure, we outline the position of the disk as seen in polarized intensity at $H$-band (see
Fig.~\ref{fig:hd142527_fig3}, top left) with black contours. We see that at the position of the two nulls the polarized color is \textit{redder} than in the rest of the disk.
In particular, there is a difference of $\approx0.5$ magnitudes between the northern and the southern null,
with the southern null being the redder one.

\subsection{Spiral Features}

\begin{figure}[ht]
  \center
   \includegraphics[width = 1\linewidth,trim = 0 0 10 25]{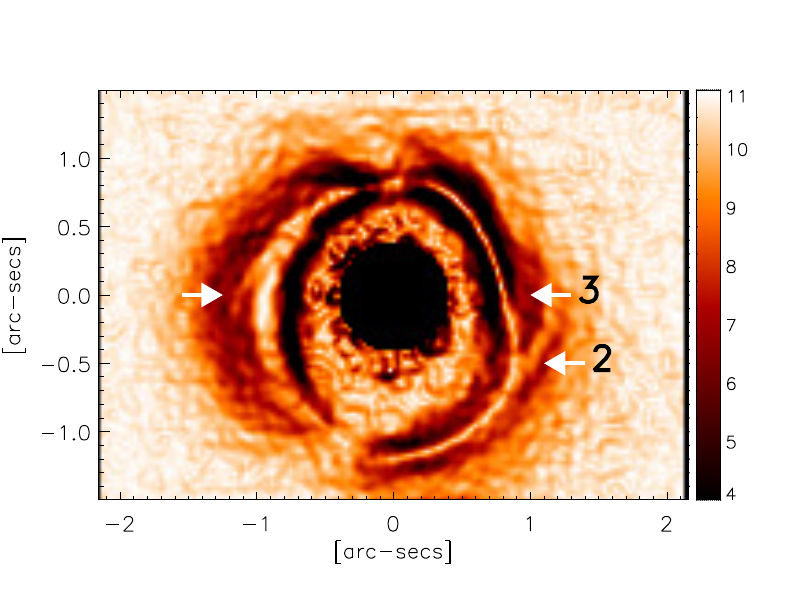}
   \caption{$P_I$ images at H-band after applying a Sobel-filter to highlight 
   the regions of the image with edges. The three arrows indicate spiral arms on the disk. 
   For comparison, we use the same notation as in \citet{2012ApJ...754L..31C} \citep[see also][]{2012A&A...546A..24R},
   to label the spiral arms. The spiral arm labelled as ``2" perfectly matches its equivalent in \citet{2012ApJ...754L..31C}.
   The spiral arm labelled as ``3" is barely visible in our images. The spiral arm on the East side is detected for the first
   time. Color bar is in arbitrary units.}
  \label{fig:hd142527_fig6}
\end{figure}

We have used different edge-filter operators over the $P_I$ images in order to bring out morphological features of the disk
such as spiral features. In Fig.~\ref{fig:hd142527_fig6} we show the result of applying a Sobel filter
\citep[or operator, see][]{2002dip..book.....G} to the $P_I$ image at $H$-band (with the inner regions of the image masked out to remove artifacts).
This filter highlights regions of the image with strong gradients. High values of the gradient magnitude correspond to places where there is rapid
change in the image values and vice versa. In order to reduce the effects of noise, the total polarized intensity image was convolved with a two dimensional
Gaussian kernel with $\sigma=1$ pixel before applying the Sobel operator. Both $H$ and $Ks$ band clearly show the spiral features labelled as ``2'' in Fig. 2 of 
\cite{2012ApJ...754L..31C} \citep[see also][]{2012A&A...546A..24R}, as well as a faint, tentative spiral feature that matches the one labelled as ``3'' on that figure.
We also clearly detect a new spiral feature on the East side, at a position angle of $PA\approx 60^{\circ}$ (see Fig.~\ref{fig:hd142527_fig6}).
These spiral features also are visible when using different edge-detection filters, so we conclude that the new spiral on the eastern side is
real, and not an artifact.

\subsection{Polarization Degree and Azimuthal Variations}

\begin{figure}[ht]
  \center
   \includegraphics[width = 1\linewidth,trim = 30 05 55 20]{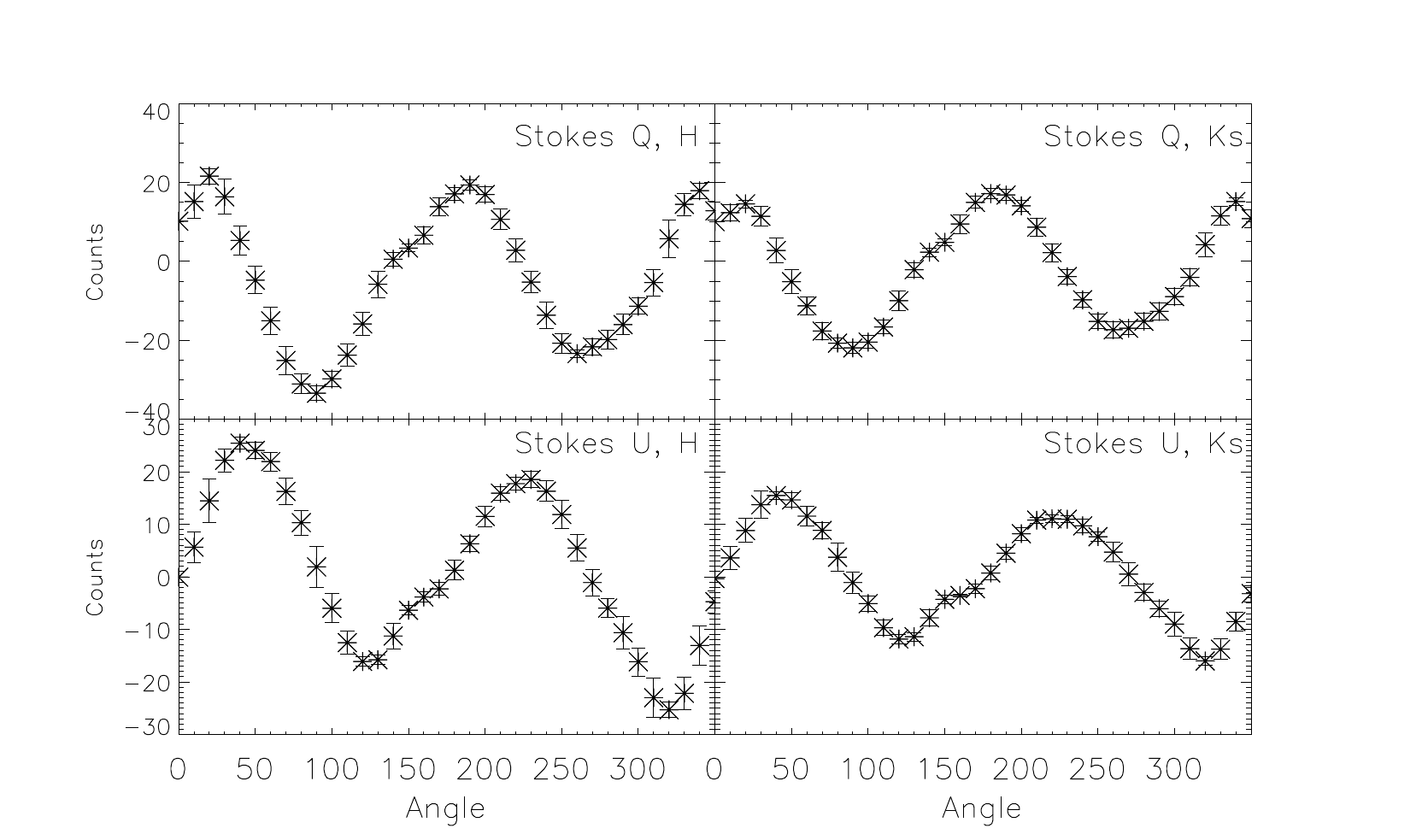}
   \caption{Azimuthal (angles computed East of North) variation of the Stokes $Q$ (top row) and $U$ (bottom row) images, for
   the $H$-band (left column) and $Ks$-band (right column) measurements.}
  \label{fig:hd142527_fig7}
\end{figure}

\begin{figure}[ht]
  \center
   \includegraphics[width = 1\linewidth,trim = 30 05 55 20]{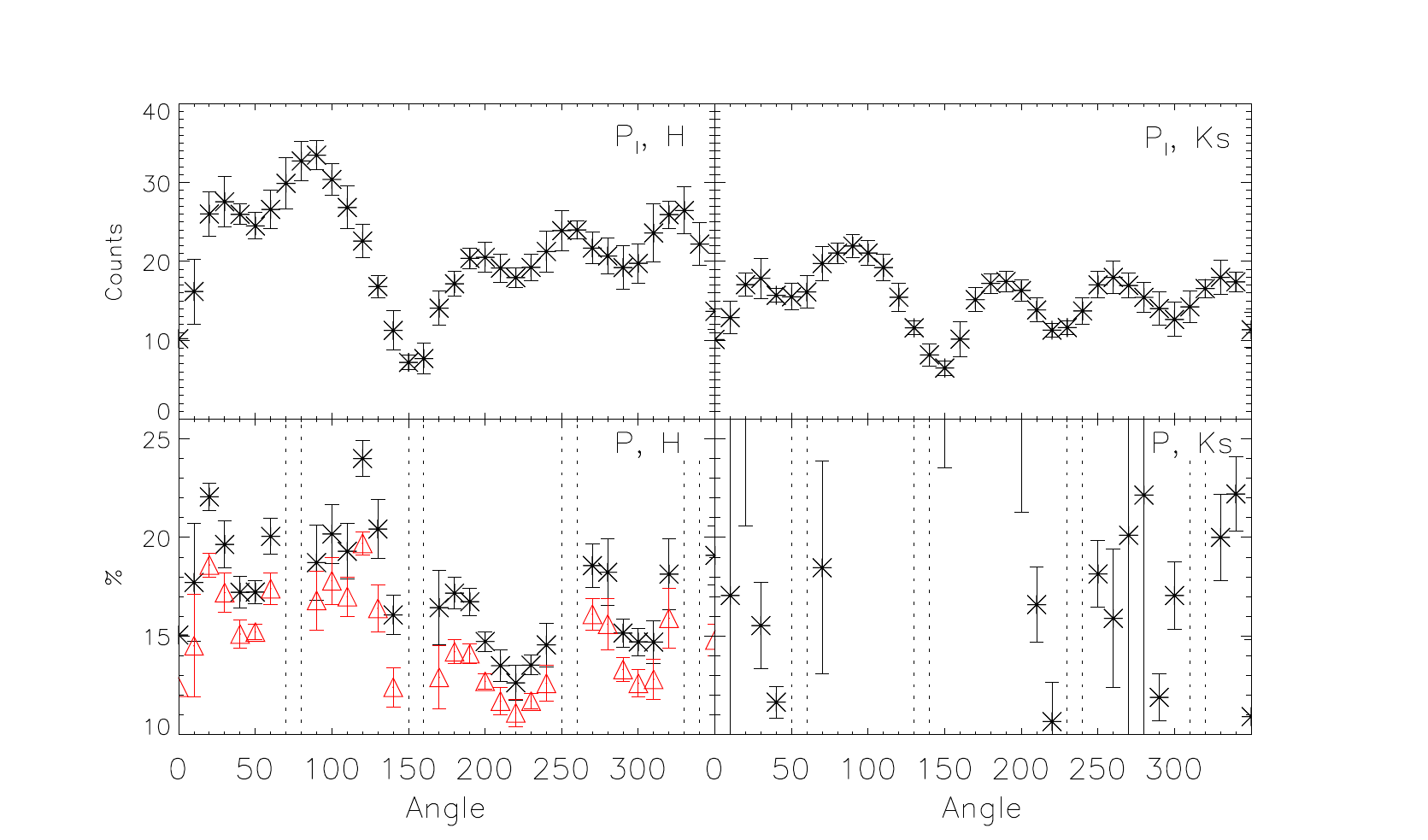}
   \caption{Azimuthal (angles computed East of North) variation of the $P_I$ (top row) and $P$ (bottom row) images, for
   the $H$ (left column) and $K$s bands (right column). The position of the telescope' spiders is indicated by the dashed lines. 
   In the $P$ plot at $H$-band, the stars and the triangles represent the lower and upper limits of $P$ (see text for a detailed explanation).
   The $P$ image at $Ks$ band is too contaminated by artifacts (see Fig.\ref{fig:hd142527_fig3}, right-bottom plot).}
  \label{fig:hd142527_fig8}
\end{figure}

\begin{figure}[ht]
  \center
   \includegraphics[width = 1\linewidth,trim = 30 05 20 20]{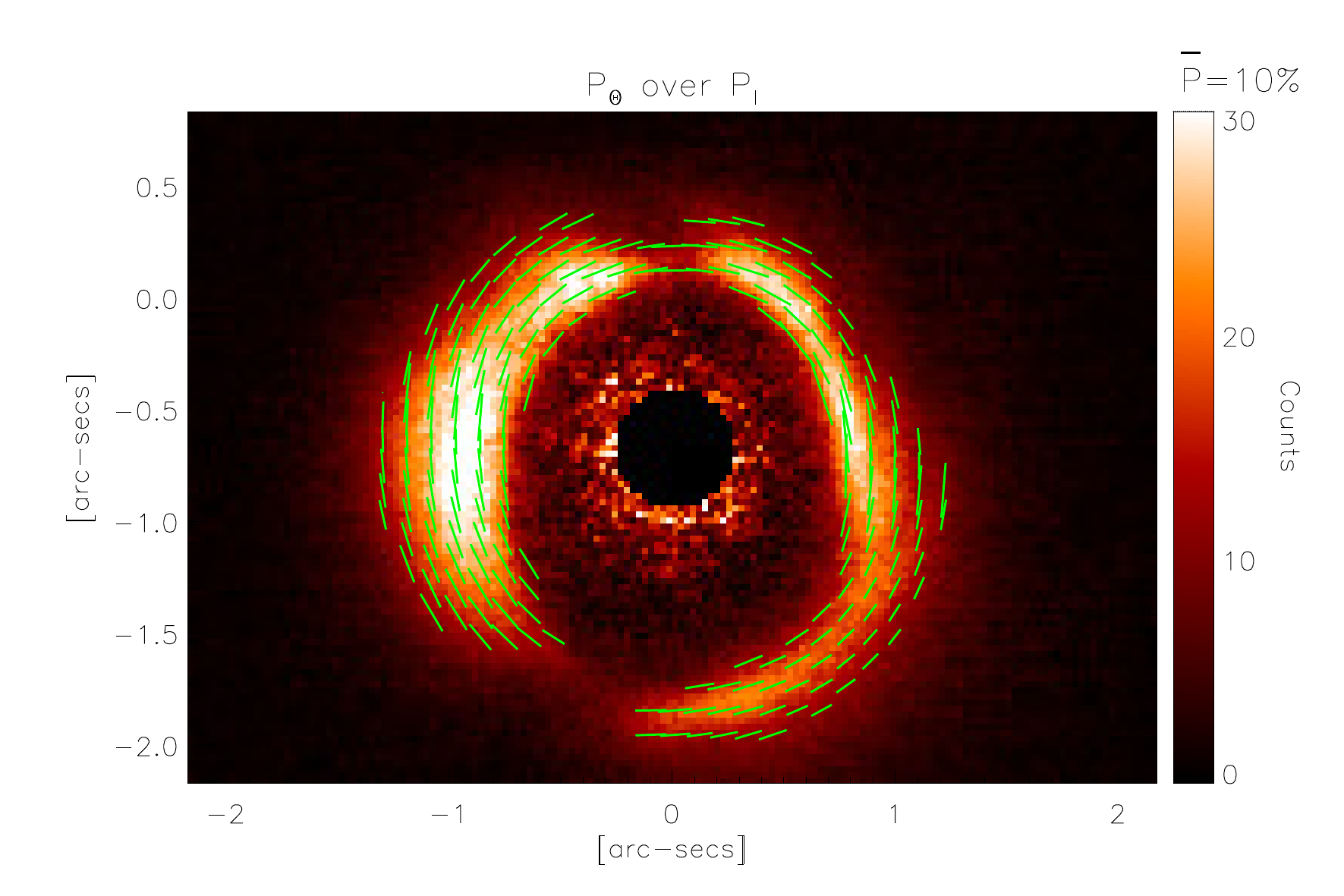}
   \caption{Polarization angle ($P_{\theta}$) indicated by the green vectors, 
   plotted over the $P_I$ image. The length of the vectors is proportional to the local polarization degree ($P$).
   The vectors are plotted in regions of the image where the polarized intensity $P_I$ }
  \label{fig:hd142527_fig9}
\end{figure}

As explained in Sect. 3.3, our $I_\mathrm{disk}$ image provide us with lower limit of the real intensity of the disk,
although we can use the sky-regions to estimate the upper limits. This translates into lower an upper
limits to the true polarization degree of the disk. We have used a 2 pixel-thick ellipsoid
divided into 36 $10^{\circ}$-sectors to compute the azimuthal variation of the $I$, $Q$ and $U$ images.
We have computed the average values of these quantities inside each sector, from which we derive the associated $P_I$ and $P$
as explained in Sect. 3. For the ellipsoid, we used the same parameters than those derived by
\citet{2012ApJ...754L..31C} to fit the disk of HD142527. The azimuthal profiles are plotted on Fig.~\ref{fig:hd142527_fig7} and Fig.~\ref{fig:hd142527_fig8}.
In all the plots, each point represents the mean value of the plotted image over the corresponding sector.
The data points that are at the position of the telescope' spiders have been removed from the $P$ plots,
since those points were severely affected by noise. The error bars represent the standard deviation inside each sector,
divided by the total amount of images used to produce the final images. 
We show the $P$ plot at $Ks$ band for completeness, but it is too contaminated by the
instrumental artifacts to be useful to derive any physical parameters, and we focus on the $P$ image at $H$-band.
The (black) stars and the (red) triangles represent the upper and lower limits of $P$, respectively.
On average, the East side of the disk has higher polarization degree ($P\lesssim19.0\%$)
than the West side ($P\lesssim15.8\%$). The polarization angle ($P_\theta$) is shown in Fig.~\ref{fig:hd142527_fig9}. 
To obtain $P_\theta$ we first bin the Stokes $Q$ and $U$ images using a binning size of 5 px.
We then compute the average $I$, $Q$ and $U$ inside each bin, and then derive the corresponding
$P_I$, $P$ and $P_\theta$. Only regions of the image where $P_I \geq 3\sigma_{P_I}$, where $\sigma_{P_I}$ is the noise of the
$P_I$ image, are considered to compute $P_\theta$. The length of the vectors is proportional to
the average of the local (upper limit of) $P$.  The centrosymmetric pattern indicates that the central star
is the source of the scattered light in our images.

\section{Discussion}

\subsection{Orientation and surface brightness of the Disk}
\label{subsec:orient}
From MIR imaging, previous studies have inferred that the disk around HD\,142527 is inclined by $\approx 20^{\circ}$, with the Western
side being the near side \citep{2006ApJ...644L.133F, 2011A&A...528A..91V}. Our observations are in good agreement
with this orientation. On the Eastern side, the inner disk wall is directly exposed to the observer and the larger (apparent) radial extension of the disk
is a natural consequence of the large vertical extent of the disk inner wall in that configuration. The large vertical extent has been suggested to
explain the very bright Mid and Far infrared excess emission \citep{2011A&A...528A..91V}.

Previous studies have estimated the NIR surface brightness of the disk along cuts at PA = 47 and 60 degrees, i.e., more or less along the North-East to
South-West direction \citep{2006ApJ...636L.153F, 2012A&A...546A..24R}. Because the disk is elliptical and not centered, these radial surface brightness
profiles peak at different radii, closer to the center in the NE direction, and farther out to the SW. Our data show the same behavior. To extract the azimuthal
SB profile in the $H$-band, we fitted ellipses to the disk along the ridge of maximum brightness. The Peak Surface Brightness is similar on both sides of the disk
in $H$-band. The $Ks$-band intensity map is too noisy to extract a reliable azimuthal profile.

\subsection{Polarization: Comparison with other disks}

Maps of the polarized intensity exist for several disks, but to our knowledge there are just four protoplanetary disks with \textit{spatially resolved} maps of the polarization
degree: GG~Tau \citep{2000ApJ...536L..89S}, AB~Aurigae \citep{2009ApJ...707L.132P} and the edge-on disk PDS 144N \citep{2006ApJ...645.1272P}, all observed with
the Hubble Space Telescope, and UX~Tau~A \citep{2012PASJ...64..124T}, observed with the {\sc subaru} telescope. PDS 144N is the only edge-on disk in this sample.
Therefore, we focus on the other three disks. In GG~Tau and AB~Aur the reported polarization
levels oscillate between a minimum of $\approx20\%$ and a maximum of $\approx50\%$, at  $2 \mu$m for AB~Aur and $1 \mu$m for GG~Tau. UX~Tau~A shows a larger
range of polarization levels, with $P$ varying between $\approx1.6\%$ and $\approx66\%$.  For all three of these disks as well as for  HD\,142527,  the polarization level
($P$) is larger on the back side of the disk, farther away from the observer, where backscattering is occurring  (assuming that the Eastern side is the back side for HD\,142527,
see Sect.~\ref{subsec:orient}).

Several factors affect the maximum polarization observed. The inclination of the disk and its flaring (i.e., opening angle), as well as the dust properties are expected to have a large
effect. HD~142527 is seen with an inclination of about $20^{\circ}$. This is a little more pole-on than the other three. The disk of AB~Aur is tilted in the range
$22^{\circ} < i < 35^{\circ}$, for GG Tau the inclination of the circumbinary ring is $\approx37^{\circ}$, and the inclination derived for UX Tau A is $46^{\circ}$. 
If we assume that the four disks have similar opening angles (i.e., flaring), then HD\,142527 is the one where scattering should on average be closest to $90\deg$.
As a consequence of this, HD\,142527 should show less differences between the front and back side because the scattering angles are also limited to a smaller range.

The range of variation of the observed polarization is indeed small, from $\approx10\%$ to $\approx25\%$, as expected for a system near pole-on. The azimuthal profiles of the 
Stokes Q and U parameters are presented in Fig.~\ref{fig:hd142527_fig7} and the azimuthal profiles of the polarization levels (and polarized intensities) are 
shown in Fig.~\ref{fig:hd142527_fig8}. Interestingly however, the polarization levels in the disk of HD~142527 are also the ones with the lowest observed
maximum (and average), although scattering at angles close to 90 degrees, as is the case here, are usually favorable to produce large polarizations.
It is tempting to attribute this behavior, the low polarization levels, to the dust properties. 

Other protoplanetary disks for which the $H$-band \textit{integrated} polarization degree is known are TW Hya
\citep{2006MNRAS.365.1348H}, HD\,100546 \citep{2011ApJ...738...23Q} and SR 21 \citep{2013ApJ...767...10F}.
In all these disks, the average $P$ does not exceeds $40\%$ at $H$ band. 

\begin{figure}[ht]
  \center
   \includegraphics[width = 1\linewidth,trim = 0 0 0 0]{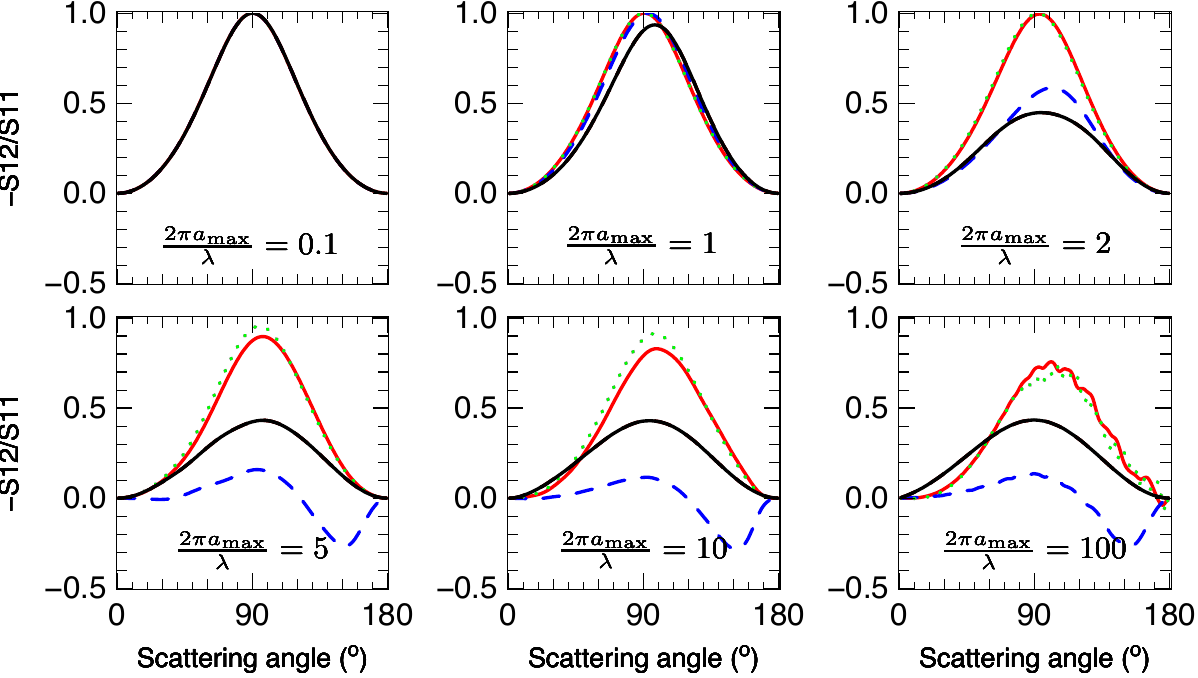}
   \caption{Plots of the Polarizability as a function of scattering angles for different mixtures of dust. The polarizability is the ratio of the -S12 over S11 elements of the
   Mueller matrix. In all case dust mixtures have power-law size distributions with $a_\mathrm{min} = 0.03 \mu m$ and a slope of $-3.5$. Six cases with increasing values of $a_\mathrm{max}$
   are considered. The red curves show the polarizability for \citet{1989ApJ...341..808M} grains (a mix of astrosilicates and amorphous carbon and $80\%$ porosity). The
   dotted green curves are for pure Astronomical Silicates with $80\%$ porosity \citep{2003ARA&amp;A..41..241D}. The blue dashed curves are for compact astronomical silicates,
   with no porosity \citep{2003ARA&amp;A..41..241D}. The black curve is for a mixture of compact, no porosity, grains made of $100\%$ amorphous carbon. All models with porosity
   have large polarizability near 90 degree scattering angles. Models with compact particles show lower maximum polarizations.}
  \label{fig:hd142527_fig10}
\end{figure}
\subsection{Dust properties}

The dust in the disk of HD\,142527 is made of a mixture of particles with different sizes and composition \citep{2011A&A...528A..91V}.
At $H$ and $Ks$ bands the dust opacity is large and the measured scattered light probes the disk surface, high above the disk mid-plane. 
A mixture of dust containing only small particles in the Rayleigh regime, i.e., with radii much smaller than the wavelength, produces isotropic
scattering and very high levels of polarization. The observed azimuthal variations of the brightness (the phase function) is compatible with nearly isotropic
scattering (peak brightnesses are equal on the back and front side). However, the observed polarization levels are significantly too low (see Fig.~\ref{fig:hd142527_fig10},
top left panel). At the surface of the disk, a mixture of dust containing only small grains can probably be ruled out.

Previous measurements at NIR and MIR were used to report particle sizes of the order of a micrometer or more
\citep{2006ApJ...636L.153F, 2006ApJ...644L.133F}. The strong emission associated with hot silicate particles
\citep{2004Natur.432..479V} is well fitted by mixtures dominated by dust grains with sizes of  $\approx1.5 \mu$m
at the rim and at the surface, from where the emission is coming from. These particles are not in the Rayleigh 
regime at $H$ and $Ks$-bands and are compatible with the observed polarization presented here.

Assuming further that the size distribution follows a power-law (e.g., with a slope of -3.5 as in previous studies), calculations also show that mixtures of particles
with significant porosity ($50\%$ or more) produce very large polarizations in all cases, with $P$ = 50\% or more for scattering angles near 90 degrees (see  Fig.~\ref{fig:hd142527_fig10}).
Mixtures with grains having large porosities are likely incompatible with the data as well. This statement is valid whether the mixture is of pure silicate or
whether it contains amorphous carbon as well. 

Distribution of compact particles where the maximum size is of order 1.0 micron or more, but with no porosity, naturally produce low levels of polarizations,
in agreement with what is observed in HD\,142527. The shape of the particles may also affect the polarization levels. We have neglected the effect of shape in our study.

\subsection{Intensity nulls}

There are two remarkable features in the disk that can be found due North and South-South-East. At these two positions the polarized intensity shows a deep minimum,
but does not go to zero  (See Fig.~\ref{fig:hd142527_fig3} and  Fig.~\ref{fig:hd142527_fig9}). These two ``nulls" are also detectable in the intensity images
(See Fig.~\ref{fig:hd142527_fig2} and  Fig.~\ref{fig:hd142527_fig3}). The polarization levels also show a marginally significant decrease at the position of the
nulls but these variations are likely part of a larger azimuthal variation pattern where the front (back) face of the disk exhibits a lower(larger) polarization. 

These nulls in HD\,142527 were previously observed in scattered light intensity by \citet{2012ApJ...754L..31C} at $K$ and $L$ bands and by \citet{2012A&A...546A..24R}
at $L$-band. They are also visible in the $Q$-band (i.e., in thermal emission at ~19 microns) in the VISIR image presented by \citet{2011A&A...528A..91V}. 

A similar ``gap" in the map of polarized intensity in the disk of AB~Aur was observed by \citet{2008ApJ...679.1574O} However, \citet{2009ApJ...707L.132P} subsequently
showed that this feature in AB~Aur is not caused by a decrease in local density. They showed instead that it is caused by a local variation of the polarization in the back side
of that disk rather than a true decrease of the surface brightness level. In HD\,142527, contrarily to AB Aur, two nulls are detected instead of one and they are located along
the major axis of the disk rather than on the back side. Because the nulls are also visible in intensity images, both in scattered light and in thermal emission, the evidence is
strong that they are tracing density and/or structural features in the disk.

The origin of these nulls is not known in details. They may be the result of surface features in the disks, such as local enhancements of the scale
height that may change the illumination pattern. In this case the nulls would trace more shadowed areas, being colder (less thermal emission) and scattering less light. It is worth noting
that the two nulls are located between spiral features identified in \citet{2012ApJ...754L..31C} and these spirals may also correspond with local changes in the density, scale height, or
at least position of the optically thick surface of the disk. In their Figure 2, the Northern null (feature  No.1 in green) is located between spiral features No. 1 and 4 (shown in red). 
The SSE null (No. 2 in green) is located between spiral features No. 1 and 2 (in red again). These spiral features could be linked to deformations on the surface of the disk, producing the
observed nulls.

Another possibility can be derived from  the more recent, high resolution sub-millimeter images of HD\,142527 that became available with ALMA. In the Northern
direction, a very conspicuous surface brightness enhancement is visible, in the form of a ``horse-shoe" or ``banana-like" feature. The observed feature is described in details in
\citet{2013Natur.493..191C}. Retrospectively, the feature could also be seen in previous SMA data, but with a lower contrast \citep{2008Ap&amp;SS.313..101O}. 
Horseshoe-like features at mm-wavelengths have been previously observed in the transition disks LkHa 330, SR 21N and HD135344B by \citet{2009ApJ...704..496B},
and very recently in IRS 48 by \citet{2013Sci...340.1199V}, and they
now seem to be a common feature of transition disks when observed with ALMA. They have been, tentatively, interpreted in terms of dust trapping \citep[e.g., Menard et al. 2013,
in preparation,][]{2013Natur.493..191C, 2013A&amp;A...550L...8B, 2013A&amp;A...553L...3A, 2013Sci...340.1199V} either by a high-pressure vortex created by a Rossby Wave Instability or by resonances
with a planet located in the disk gap. For HD\,142527, the maximum intensity observed in the ``horseshoe" is nearly azimuthally coincident with the Northern null. It is tempting to relate the two features.
The enhanced dust density and/or abundance of large grains in the vortex (the horseshoe) may lead to less scattered light from the surface because, if transported close to the surface, grains
much larger than the wavelength are less efficient scatterers and polarize differently than small grains. Similarly, the larger dust density in the vortex may lead to a lower local dust temperature,
resulting in less thermal emission at 20 microns ($Q$-band). At the same time, the presence of more dust, in particular larger dust more prone to be trapped by the gas high-pressure, 
would naturally produce the increase in sub-millimeter emission, as observed in the ALMA images. The other null, located SSE, is more difficult to interpret in that context. These suggestions
will be tested in more details in a forthcoming paper. 

A third possibility, obvious and very popular these days, is that the nulls are the traces of hidden and bigger bodies that are currently evacuating the disk material around them, e.g.,
planets in the runaway accreting phase. This possibility seems difficult to reconcile with all the observations however, in particular the difference in sub-millimeter properties between
the two nulls.

\section{Summary and Conclusions}

We present $H$ and $Ks$ imaging polarimetry of HD\,142527. We use our polarized intensity ($P_I$)
images to morphologically describe the disk. The presence of two ``nulls" or regions with lower
$P_I$ at the North (top null) and South-East (bottom null) is clearly detected. Our results confirm
previous observations of this system, showing a heavily dust-depleted inner gap extending
up to 130 AU. We do not detect the continuum counterpart of the HCO+ streamers 
claimed by \citet{2013Natur.493..191C}, but our sensitivity to faint extended features is limited inside the gap.
We detect a new spiral feature on the East side of the disk by applying an edge-detection
filter to our polarized intensity images. We obtain a direct image of the disk at $H$-band by applying classical PSF-subtraction.
This image also shows the presence of the two nulls, which rules out geometrical scattering effects as a possible
explanation for them. We combine the $P_I$ and the PSF-subtracted images to place
upper and lower limits to the polarization degree in this disk, which is found to vary between $10\%$ and $25\%$.
In both cases, the Eastern side of the disk shows highest polarization degree than the Western side.

HD\,142527 shows lower polarization degree and lower azimuthal variations than the protoplanetary
disks GG Tau, AB Aurigae and UX Tau A. The small azimuthal variation of the polarization degree
is expected in case the disk is seeing nearly pole-on, which is in agreement with previous results
suggesting an inclination of $\approx20$ degrees for HD\,142527. Furthermore, the low polarization
degree indicates that the distribution of dust particles on the disk's surface must extend at least to
micron sizes or more. Comparison with different grain populations suggests that the dust grains on the surface of the disk
are not very porous, pointing towards compact grains. 
We also compare our images with ALMA Band 7 continuum observations of the same disk. These images show
that mm-sized dust follows a horse-shoe shape, with its maximum emission placed at roughly the same
azimuthal position as the Northern null observed on our images. We have suggested possible explanations for the nulls,
such as Rossby Wave Instabilities or local enhancements of the scale height of the disk. It is also tentative to link
these nulls with the formation of planetesimals, although more evidence is needed to support this hypothesis.

	
\begin{acknowledgements}
	We are Grateful to the ESO staff, and Julien Girard in particular, for their help during the observations.
	This research was funded by Millenium Science Initiative, Chilean Ministry of Economy,
	Nucleus P10-022-F. AJ also acknowledges support from Fondecyt project 1130857.
	We acknowledge funding from the European Commission's 7$^\mathrm{th}$ Framework Program
	contract PERG06-GA-2009-256513) and from Agence Nationale pour la Recherche (ANR) of France
	under contract ANR-2010-JCJC-0504-01.
\end{acknowledgements}

\bibliographystyle{aa.bst}	
\bibliography{biblio}		

\clearpage
\newpage
\end{document}